\documentclass[a4paper,11pt]{article}
\usepackage{pos}
\usepackage{overpic}
\newcommand{\dif}{\mathrm{d}}
\title{Exploration of Two-Photon Exchange in proton Form Factors at BESIII}
\author{Lei Xia}

\affiliation[a]{University of Science and Technology of China}

\emailAdd{xial@ustc.edu.cn}

\abstract{The proton and neutron make up over 99.9\% of visible matter in the universe. The internal structure of protons, governed by ElectroMagnetic Form Factors (EMFFs), has been probed in both space-like (SL) and time-like (TL) regions. The BESIII experiment has achieved the most precise TL proton FF measurements to date, providing new insights into proton structure. However, higher-order effects in EMFFs, especially the impact of Two-Photon Exchange (TPE), remain poorly understood due to experimental challenges in $e^{+}e^{-}$ annihilation. In this work, we analyze high-statistics BESIII data across a wide c.m. energy range and observe, for the first time, a significant angular asymmetry (8.6$\sigma$) in proton production. We demonstrate that this asymmetry arises from TPE, using a Born-like event selection, and extract the corresponding TPE-related EMFF properties. The observed interference between two-photon and one-photon exchanges (OPE$\otimes$TPE) is consistent with expectations at the level of the fine-structure constant relative to the OPE.}

\FullConference{
}

\begin{document}
\maketitle

\section{Introduction}

Our understanding of matter has evolved from molecules down to the inner structure of nucleons. The proton, discovered by Rutherford in 1919, together with the neutron, forms the nucleus of atoms. Early observations, such as Stern's measurement of anomalous nucleon magnetic moments~\cite{Stern1}, hinted at substructure, which Hofstadter’s electron scattering experiments in the 1950s~\cite{Hofstadter1,McAllister1} confirmed. Later, deep inelastic scattering experiments by Friedman, Kendall, and Taylor led to the quark model~\cite{Bloom1,Breidenbach1}. Key to unraveling the proton’s structure is understanding the distributions of its electric charge and magnetization.

Historically, limited statistics in EMFF studies constrained knowledge to the OPE approximation. In recent decades, polarization techniques in $ep$ scattering have revealed discrepancies in the ratio $\mu G_{E}/G_{M}$ between unpolarized and polarized measurements, now known as the “proton form factor puzzle”~\cite{Zhou1}. This drove a renewed focus on the TPE effect as a higher-order correction.
Direct measurements of TPE in the SL region, such as those at VEPP-3~\cite{VEPP-3_1,VEPP-3_2}, DORIS/OLYMPUS~\cite{OLYMPUS1}, and JLab (CLAS, Hall A)\cite{CLAS1,CLAS2,JLAB1}, and MAMI\cite{MAMI1}, consistently find TPE contributions, particularly at higher $Q^2$. However, these experiments often lack full kinematic coverage in the critical $Q^2=2$-$7$~GeV$^2$ region where the $\mu G_{E}/G_{M}$ discrepancy is largest. Altogether, these studies point to important higher-order TPE effects in understanding the proton's electromagnetic structure\cite{Carlson1,Arrington1}.

The electromagnetic form factor $F_{3}$ becomes significant within the framework of TPE processes, where it encodes higher-order spin-flip contributions to the proton current. In the differential cross section, interference terms involving $F_{3}^{*}$ and the leading form factors $G_{E}$ and $G_{M}$ generate angular distribution components that are odd in $\cos\theta$. While $F_{3}$ itself is not directly accessible, its presence can be inferred through observable forward-backward or charge asymmetries in the proton’s angular distribution, thereby providing experimental sensitivity to $\operatorname{Im}F_3$.
In the SL region, such sensitivity has been achieved through measurements of transverse single-spin asymmetries, most notably by the MAMI experiment~\cite{MAMI1}. In contrast, our study delivers the first opportunity to probe $\operatorname{Im}F_3$ in the TL regine, using angular observables alone. Unlike SL elastic scattering, where imaginary parts of the amplitudes are typically suppressed, the complex-valued nature of the electromagnetic form factors in the TL region makes the impact of TPE more pronounced and easier to detect. This methodology leverages both the enhancement of signal amplitudes and a more direct sensitivity to TPE dynamics.

\section{Charge Asymmertry}
\subsection{Angular distribution analysis}

The observed asymmetry can be characterized using the following observable:
\begin{eqnarray}
\label{AC}
\frac{\dif A_{C}}{\dif\cos\theta} = \frac{\dif \sigma_{p}(\cos\theta) - \dif \sigma_{\bar{p}}(\cos\theta)}{\dif \sigma_{p}(\cos\theta) + \dif \sigma_{\bar{p}}(\cos\theta)},
\end{eqnarray}
where $\frac{\dif A_{C}}{\dif\cos\theta}$ is computed by comparing the angular distributions of both protons and antiprotons. Here, $\dif \sigma_{\bar{p}}(\cos\theta)$ refers to the differential cross section for $e^{+}e^{-}\to p\bar{p}$, determined from the fully corrected antiproton angular distribution.

\begin{figure}[htbp]
\begin{center}
\centering
\vskip+20pt
\mbox{
\hskip-6pt
  \begin{overpic}[width=6.7cm,height=5.0cm,angle=0]{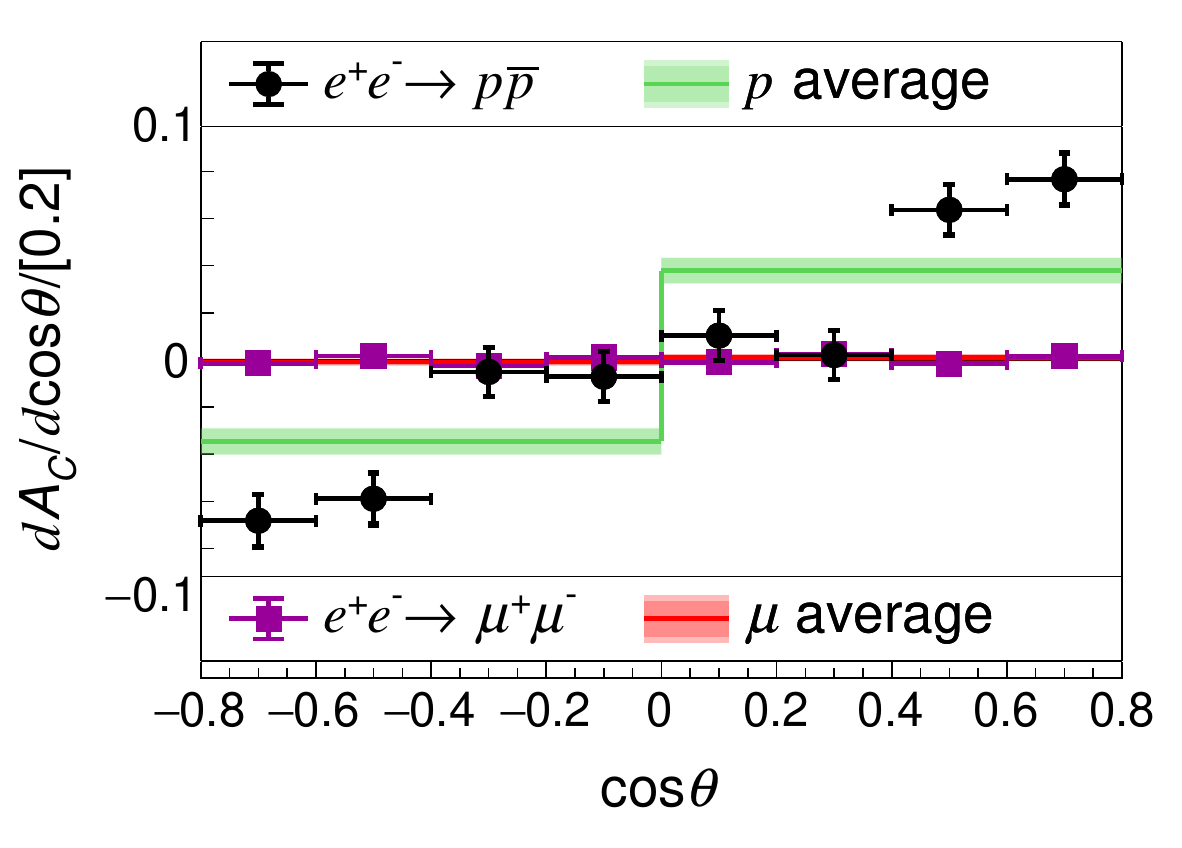}
  \put(0,70){$(a)$}
  \end{overpic}
  \hskip-3.9pt
   \begin{overpic}[width=6.7cm,height=5.0cm,angle=0]{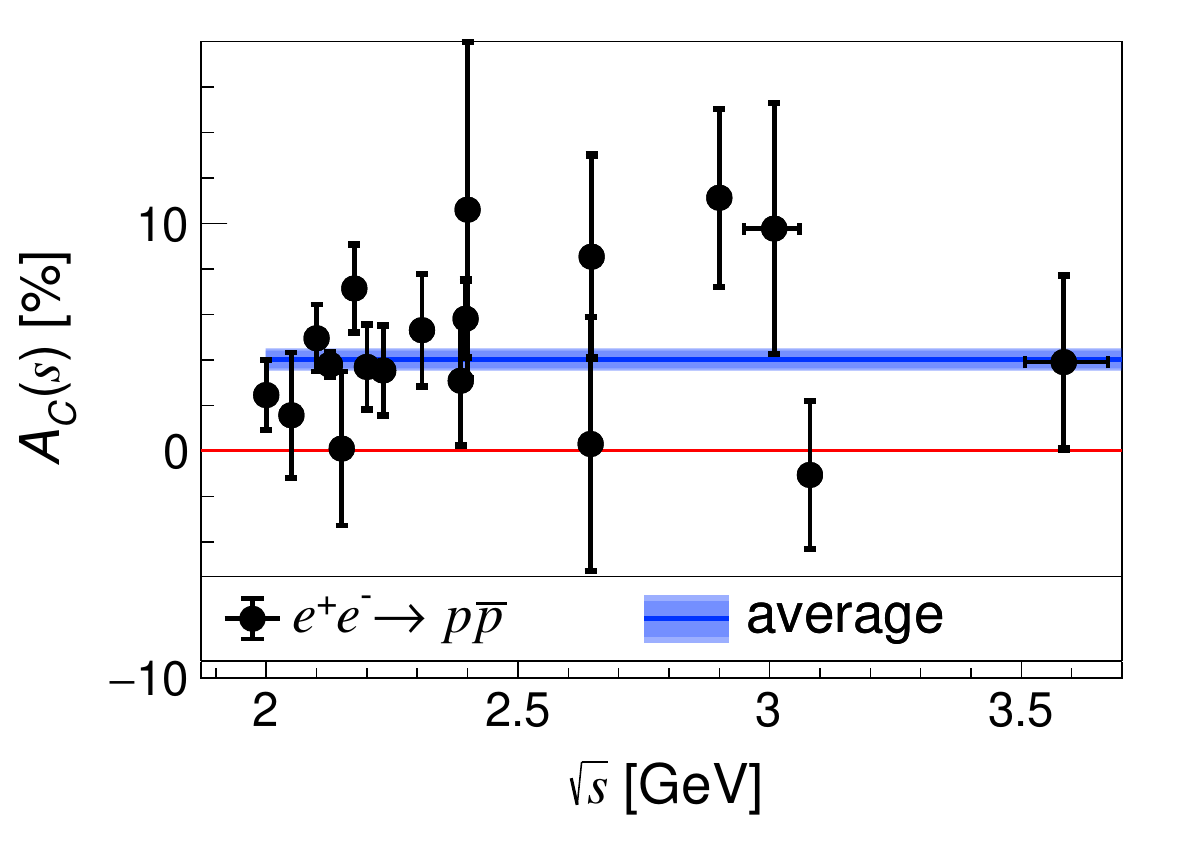}
  \put(0,70){$(b)$}
  \end{overpic}  
}
 \caption{(a) The charge asymmetry in the processes $e^{+}e^{-}\to p\bar{p}$ and $e^{+}e^{-}\to \mu^{+}\mu^{-}$ is depicted by the black dots with error bars and the purple squares with error bars, respectively, with the latter included for comparison. The overall charge asymmetry is represented by a green and red solid line with band, and a red solid line with band at $\sqrt{s}=2.125$~GeV. Both statistical and systematic uncertainties are included in this figure.
 (b) The overall dependence of $A_{c}$ on $\sqrt{s}$ for the process $e^{+}e^{-}\to p\bar{p}$ (represented by the black dots with error bars) is analyzed and the average is shown in a blue line and band. Both statistical and systematic uncertainties are included in this figure.
 }
\label{fig::ACdistribution}
\end{center}
\end{figure}

At $\sqrt{s}=2.125$~GeV, the distribution of $\dif A_{C}/\dif\cos\theta$—shown as black data points with error bars in Fig.\ref{fig::ACdistribution}(a)—reveals a marked deviation from perfect symmetry. To further quantify the effect, the average charge asymmetry is extracted by integrating over the relevant regions:
\begin{eqnarray}
\label{ACA}
A_{C} = \frac{\sigma_{p}(\cos\theta>0) - \sigma_{\bar{p}}(\cos\theta>0)}{\sigma_{p}(\cos\theta>0) + \sigma_{\bar{p}}(\cos\theta>0)},
\end{eqnarray}
as illustrated by the green solid line and its associated uncertainty band. This parameter, $A_{C}$, provides a direct measurement of the asymmetry linked to the ratio $\sigma_{p\bar{p}}^{\text{OPE}\otimes\text{TPE}}/\sigma_{p\bar{p}}^{\text{OPE}}$, which is determined from model-dependent fits to the angular distributions.
Comprehensive results for all center-of-mass (c.m.) energies are presented in Fig.\ref{fig::ACdistribution}(b).

\subsection{Two Photon Exchange}

To establish a solid reference for identifying such effects, we first examine the conventional case of one-photon exchange (OPE), where the proton’s angular distribution is expected to be symmetric and is well-understood theoretically. Specifically, in the OPE framework, the EMFFs in the TL region can be determined by analyzing the differential cross section for $e^{+}e^{-} \to p\bar{p}$. Here, the polar angle $\theta$-defined as the angle between the outgoing proton (or antiproton) and the incident positron in the c.m. frame-should display a perfectly symmetric, $\cos\theta$-even distribution~\cite{Hand1}:
\begin{eqnarray}
\begin{aligned}
\label{dsigma}
\frac{\dif \sigma_{p\bar{p}}^{\text{OPE}}}{\dif\cos\theta}=a_{0}(q^{2})+a_{2}(q^{2})\cos^{2}\theta,
\end{aligned}
\end{eqnarray}
$a_{0}=\frac{\pi\alpha_{\text{EM}}^{2}\beta C}{2q^{2}}\left[|G_{M}(q^{2})|^{2}+\frac{4m_{p}^{2}c^{2}}{q^{2}}|G_{E}(q^{2})|^{2}\right]$ and $a_{2}=\frac{\pi\alpha_{\text{EM}}^{2}\beta C}{2q^{2}}\left[|G_{M}(q^{2})|^{2}-\frac{4m_{p}^{2}c^{2}}{q^{2}}|G_{E}(q^{2})|^{2}\right]$ are defined as explicit quadratic functions of $|G_{E}(q^{2})|$ and $|G_{M}(q^{2})|$, depending only on $q^{2}$. The parameter $\beta$ represents the proton velocity in the c.m. frame, $m_{p}$ is the proton mass, and $C$ is the Coulomb factor~\cite{Rinaldo1}. Recent theoretical work shows that higher-order effects like TPE are essential for a complete description of the proton’s electromagnetic structure~\cite{Gakh}, and that the traditional interpretation based solely on Eq.(\ref{dsigma}) is insufficient. In TL measurements, the $\cos\theta$ dependence of these effects reflects their $C$-odd character~\cite{dianyong}:
\begin{eqnarray}
\label{dint}
\begin{aligned}
\frac{\dif \sigma_{p\bar{p}}^{\text{int}}}{\dif\cos\theta}\approx a_{1}(q^{2})\cos\theta+a_{3}(q^{2})\cos^{3}\theta,
\end{aligned}
\end{eqnarray}
The coefficients $a_{1}(q^{2})$ and $a_{3}(q^{2})$ are observables determined experimentally from the angular distributions of final-state protons, and they reflect the asymmetries arising from OPE$\otimes$TPE interference. These coefficients are given by $a_{1}(q^{2}) = \frac{\pi\alpha_{\text{EM}}^{2}\beta C}{q^{2}} b_{1}(q^{2})$ and $a_{3}(q^{2}) = \frac{\pi\alpha_{\text{EM}}^{2}\beta C}{q^{2}} b_{3}(q^{2})$, showing that they are directly proportional to the theoretical interference terms $b_{1}(q^{2})$ and $b_{3}(q^{2})$.
The terms involving $F_{3}$ appear through products of its imaginary part, $F_{3}^{*}$, with the real parts of $G_{E}$ and $G_{M}$. Explicitly, these are expressed as: 
\footnotesize
\begin{eqnarray}
\label{b1b2}
\begin{aligned}
&b_{1}(q^{2})=\Re\left\{G_{M}\Delta g_{M1}^{*}+\frac{4m_{p}^{2}c^{2}}{q^{2}}G_{E}\Delta g_{E1}^{*}-\frac{\beta}{1-\beta^{2}}\left[G_{M}(q^{2})-\frac{4m_{p}^{2}c^{2}}{q^{2}}G_{E}(q^{2})\right]F_{3}^{*}(q^{2})\right\}\\
&b_{3}(q^{2})=\Re\left\{G_{M}(\Delta g_{M1}^{*}+\Delta g_{M3}^{*})+\frac{4m_{p}^{2}c^{2}}{q^{2}}G_{E}(\Delta g_{E3}^{*}-\Delta g_{E1}^{*})-\frac{\beta}{1-\beta^{2}}\left[ G_{M}(q^{2})-\frac{4m_{p}^{2}c^{2}}{q^{2}}G_{E}(q^{2})\right]F_{3}^{*}(q^{2})\right\}
\end{aligned}
\end{eqnarray}
\normalsize
These quantities are sensitive to the imaginary components of the electromagnetic form factors, namely $G_{E}^{}$, $G_{M}^{}$, and $F_{3}^{*}$.
These corrections are parameterized by the coefficients $\Delta g_{E1}^{}$, $\Delta g_{E3}^{}$, $\Delta g_{M1}^{}$, and $\Delta g_{M3}^{}$, which characterize the $\cos\theta$ and $\cos^{3}\theta$ dependencies of the TPE-induced modifications to $G_{E}^{}$ and $G_{M}^{}$. Specifically, the corrections are expressed as $\Delta G_{E}^{} = g_{E1}^{}\cos\theta + g_{E3}^{}\cos^{3}\theta$ and $\Delta G_{M}^{} = g_{M1}^{}\cos\theta + g_{M3}^{}\cos^{3}\theta$.
The coefficients $b_{1}$ and $b_{3}$ encapsulate the OPE$\otimes$TPE interference terms. An asymmetric model is constructed as follows:
\begin{eqnarray}
\label{dsigmaasy}
\frac{\dif \sigma_{p\bar{p}}}{\dif\cos\theta}=\frac{\dif \sigma_{p\bar{p}}^{\text{OPE}}}{\dif\cos\theta}+\frac{\dif \sigma_{p\bar{p}}^{\text{int}}}{\dif\cos\theta}\approx a_{0}(q^{2})+a_{1}(q^{2})\cos\theta+a_{2}(q^{2})\cos^{2}\theta+a_{3}(q^{2})\cos^{3}\theta,
\end{eqnarray}
In this framework, the even-order terms ($a_{0}$ and $a_{2}$) correspond to the symmetric OPE contribution, while the odd-order terms ($a_{1}$ and $a_{3}$) reflect the asymmetry from OPE$\otimes$TPE interference. The coefficients $b_{1}$ and $b_{3}$ can be extracted from the fitted $a_{1}$ and $a_{3}$ values. To date, no baryon asymmetries defined by Eq.(\ref{dint}) have been observed with significance above 8$\sigma$. BESIII’s precise TL proton form factor measurements now enable detailed studies of such asymmetries in the TL region\cite{Lei1,Lei2,Pacetti2}.

Since protons and antiprotons are produced back-to-back, their angular distributions are inherently symmetric with respect to each other. Although both particles are detected, our analysis focuses on the proton angular distribution to extract the coefficients $b_{1}$ and $b_{3}$. 
The differential cross section $\dif \sigma_{p}(\cos\theta)$ derived from the proton angular distribution is used as a representative for $\dif \sigma_{p\bar{p}}(\cos\theta)$ in the process $e^{+}e^{-}\to p\bar{p}$ at 2.125~GeV.

\begin{figure}[htbp]
\begin{center}
\centering
\vskip+20pt
\mbox{
\hskip-6pt
  \begin{overpic}[width=4.5cm,height=3.3cm,angle=0]{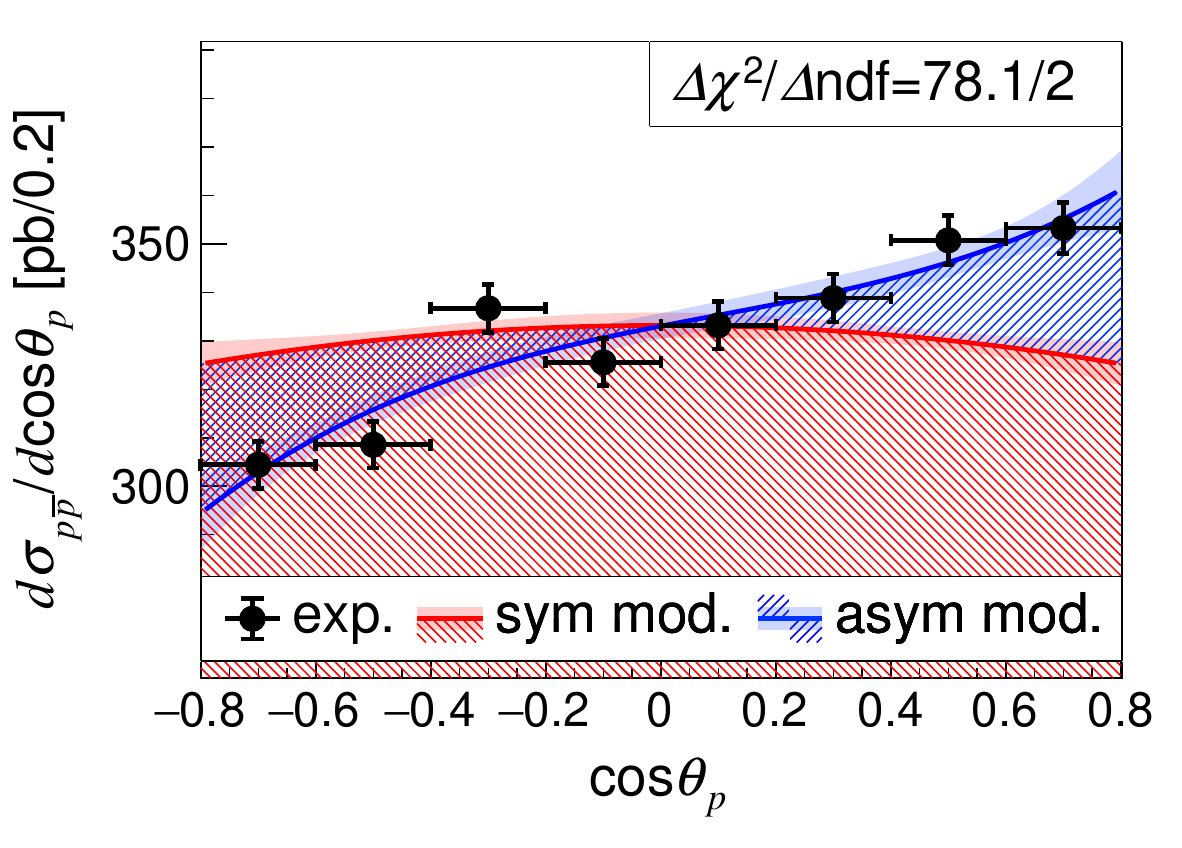}
  \put(0,70){$(a)$}
  \end{overpic}
\hskip-3.9pt
  \begin{overpic}[width=4.5cm,height=3.3cm,angle=0]{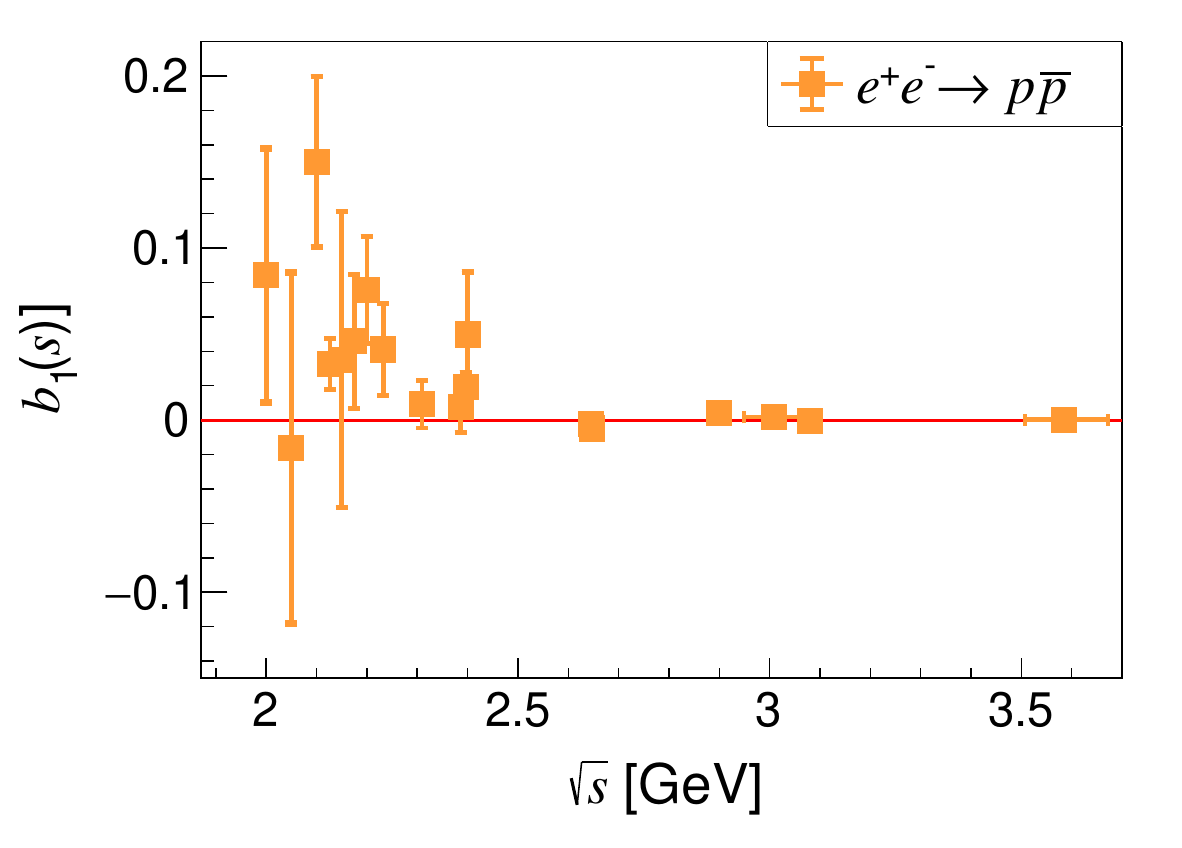}
  \put(0,70){$(b)$}
  \end{overpic}
 \hskip-3.9pt 
  \begin{overpic}[width=4.5cm,height=3.3cm,angle=0]{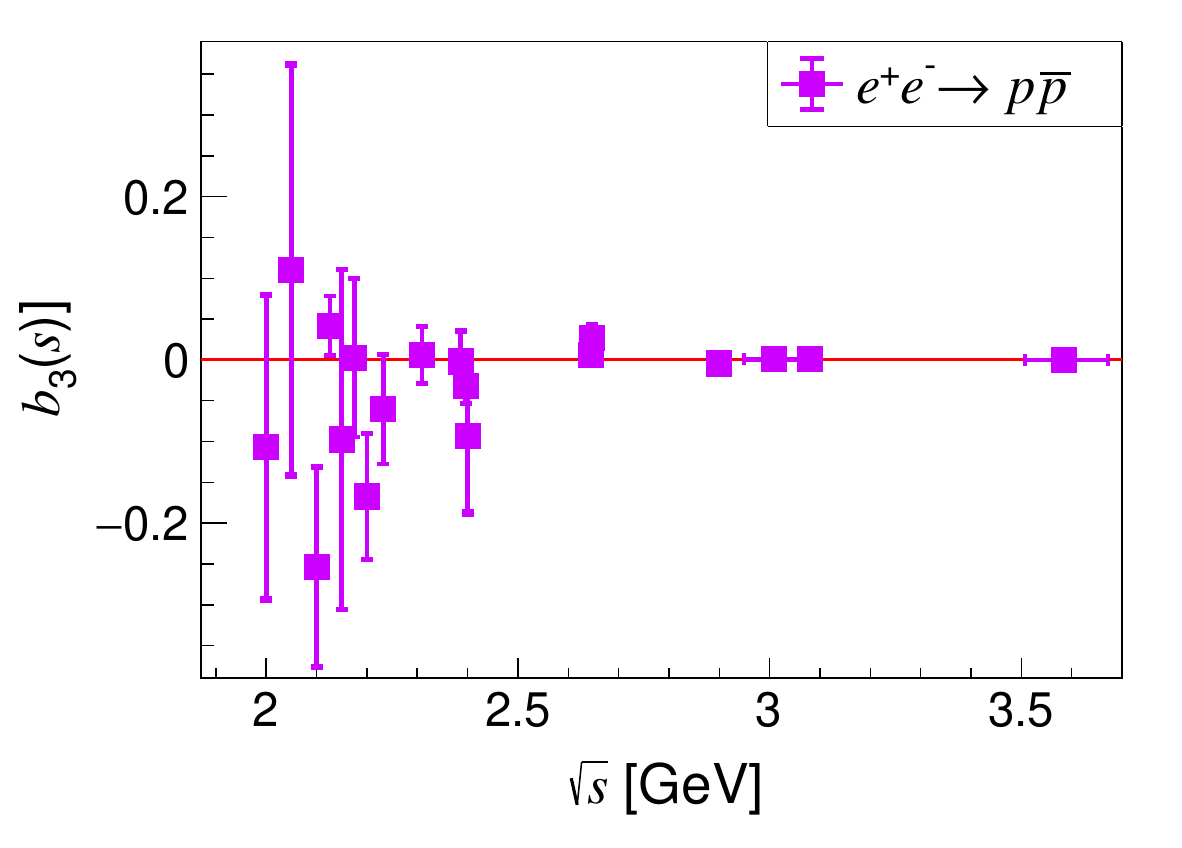}
  \put(0,70){$(c)$}
  \end{overpic}  
}
 \vskip+20.0pt
 \caption{(a) The fit to the corrected proton distribution $\dif \sigma_{p\bar{p}}/\dif\cos\theta$ at $\sqrt{s}=2.125$~GeV, using a symmetric model [illustrated by the red solid line and band, as equation (\ref{dsigma})] and an asymmetric model [illustrated by the blue solid line and band, equations(\ref{dsigma})+(\ref{dint})]. The red shaded area represents the contribution from symmetry, while the blue shaded portion indicates the asymmetry contribution. Only statistical uncertainties are shown in this figure.
 (b) The the interference coefficients $b_{1}(q^{2})$, is analyzed and represented by the orange dots with error bars. Both statistical and systematic uncertainties are included in this figure.
 (c) The the interference coefficients $b_{3}(q^{2})$, is analyzed and represented by the magenta dots with error bars. Both statistical and systematic uncertainties are included in this figure.}

\label{fig::distribution}
\end{center}
\end{figure}

Figure~\ref{fig::distribution}(a) displays the measured angular distribution at $\sqrt{s}=2.125$~GeV. A $\chi^{2}$ fit using Eq.(\ref{dsigma}) (shown as a red solid line and band) is compared to a fit including both Eq.(\ref{dsigma}) and Eq.(\ref{dint}) (depicted by the blue line and band). The latter yields a much better agreement with the data, highlighting a significant asymmetry in the angular distribution, observed with a statistical significance of 8.6$\sigma$ at 2.125~GeV. This significance is obtained from a $\Delta\chi^{2}/\Delta ndf=78.1/2$, where $ndf$ denotes the change in the number of degrees of freedom.

From the fits, the interference coefficients $b_{1}(q^{2})$ and $b_{3}(q^{2})$ are extracted, quantifying the $\cos\theta$- and $\cos^{3}\theta$-dependent terms associated with OPE$\otimes$TPE interference. At $\sqrt{s}=2.125$~GeV, the results are $b_{1}(2.125~\text{GeV})=(3.27\pm1.49\pm0.19)\times 10^{-2}$ and $b_{3}(2.125~\text{GeV})=(4.16\pm3.66\pm0.39)\times 10^{-2}$, clearly demonstrating a nonzero asymmetry and enabling the construction of the OPE$\otimes$TPE term. In the plot, the red shaded region represents the OPE contribution, while the blue region corresponds to the OPE$\otimes$TPE effect. The ratio of the asymmetric to the symmetric component, $\sigma_{p\bar{p}}^{\text{OPE}\otimes\text{TPE}}/\sigma_{p\bar{p}}^{\text{OPE}}=(3.79\pm0.53\pm0.23)\%$, is found to be of the order of the electromagnetic coupling $\alpha_{\text{EM}}$. 
Fig.~\ref{fig::distribution} (b) and (c) summarize the extracted $b_{1}(q^{2})$ and $b_{3}(q^{2})$ values at various energy levels.

\section{Radiation corrections}

\begin{figure}[htbp]
\begin{center}
\centering
\vskip+0pt
\mbox{
\hskip-6pt
  \begin{overpic}[width=6.7cm,height=5.0cm,angle=0]{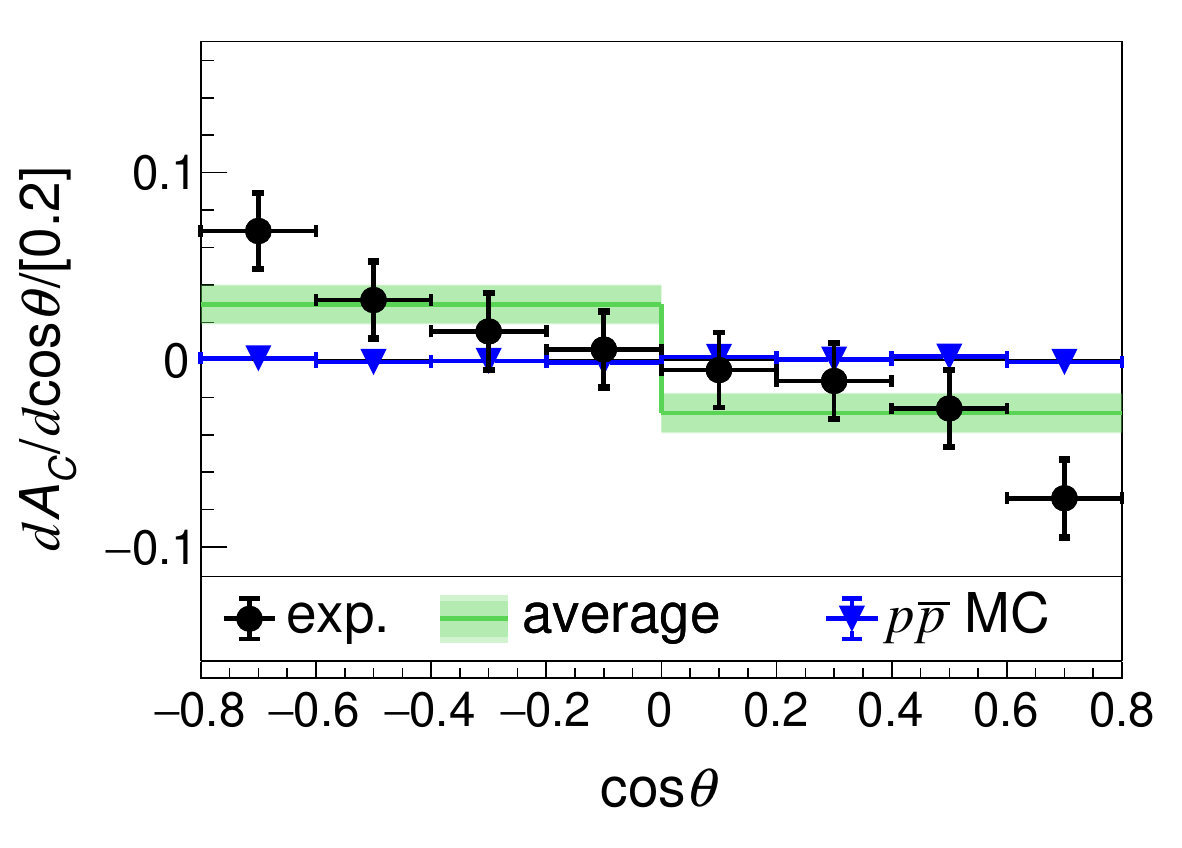}
  \put(0,70){$(a)$}
  \end{overpic}
\hskip-3.9pt
  \begin{overpic}[width=6.7cm,height=5.0cm,angle=0]{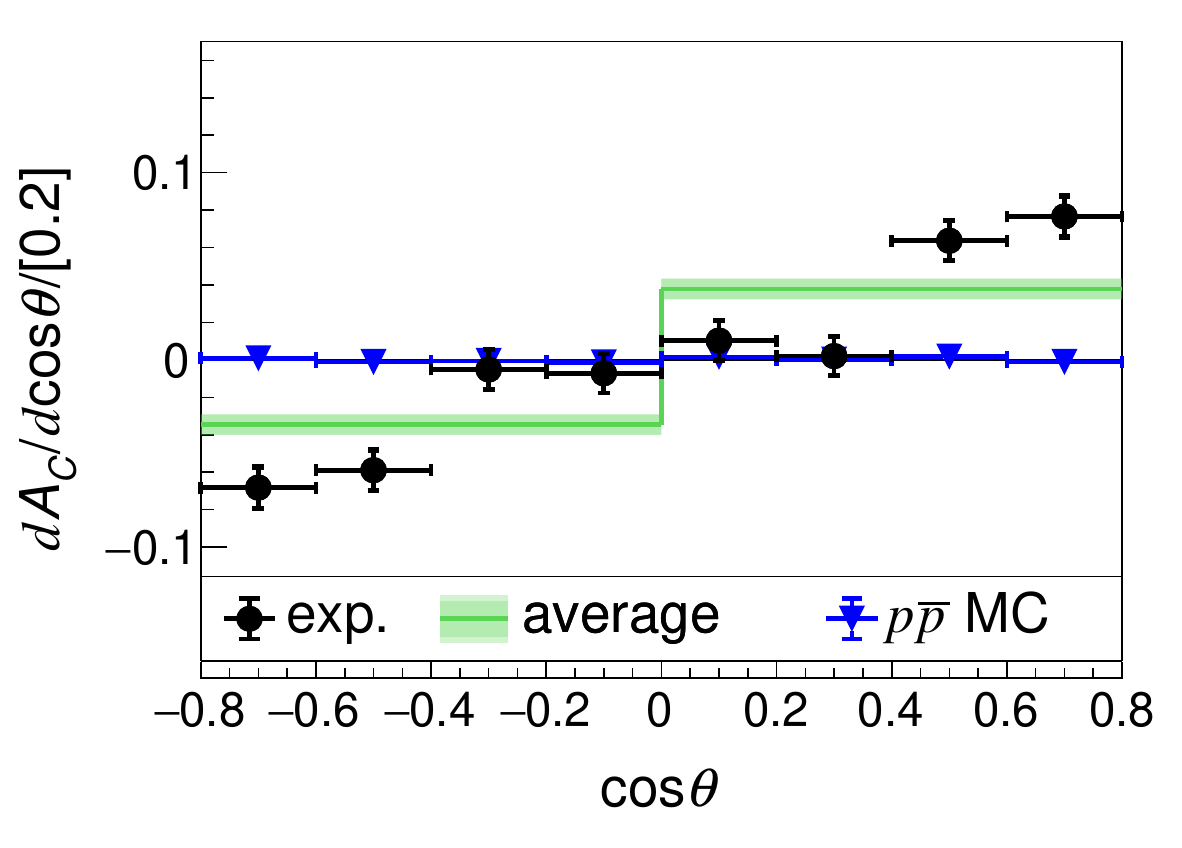}
  \put(0,70){$(b)$}
  \end{overpic}
}
 \vskip+20.0pt
 \caption{The observed charge asymmetry distributions $\dif A_{C}/\dif\cos\theta$ before introducing any explicit constraint on $M_{p\bar{p}}$, across two energy intervals: (a) 2.0900-2.1228~GeV/$c^{2}$, and (b) 2.1228-2.1499~GeV/$c^{2}$, are showcased with corrected data points (black dots with error bars) and contrasted with Monte Carlo (MC) simulations (blue triangles with error bars) that incorporate vacuum polarization (VP) and radiation effects. The overall charge asymmetry for each energy intervals, illustrated by the green solid line and band. Both statistical and systematic uncertainties are included in this figure. The results shown are based on a Born-like event selection ($M_{p\bar{p}}>2.1228$~GeV), where TPE contributions are expected to be dominant.
}
\label{fig::energyintervals}
\end{center}
\end{figure}

Strict momentum and event selection criteria largely suppress events with hard photon emission (energy >1MeV), yielding a sample dominated by soft radiation, where the OPE$\otimes$TPE effect at order $\alpha_{EM}^{3/2}$ becomes more visible. Nevertheless, soft photon events, especially those affected by initial-final state interference (IFI), can still introduce angular asymmetries\cite{phokhara}.
To determine whether a specific cut on $M_{p\bar{p}}$ (the reconstructed invariant mass of the $p\bar{p}$ pair) is needed, we analyze events after all standard selections but before imposing any explicit $M_{p\bar{p}}$ requirement. We categorize events based on $M_{p\bar{p}}$: those with lower values are identified as "radiative" (with energy loss from photon emission, mainly IFI), while those near the nominal mass are considered "Born-like" (minimal energy loss, TPE-dominated).
By comparing the charge asymmetry $\dif A_{C}/\dif\cos\theta$ across different $M_{p\bar{p}}$ regions, we can distinguish IFI from TPE contributions. As illustrated at $\sqrt{s}=2.125$~GeV in Fig.\ref{fig::energyintervals}, $\dif A_{C}/\dif\cos\theta$ is plotted for two $M_{p\bar{p}}$ intervals, separated at $M_{p\bar{p}} = 2.1228$~GeV, which corresponds to a $-0.075\sigma$ shift in the track momentum window mapped to $M_{p\bar{p}}$.

To better determine the source of the observed asymmetry, we analyze the process $e^{+}e^{-} \to \mu^{+}\mu^{-}$, which benefits from high statistics and precise QED predictions for TPE and IFI effects. Using MC efficiency corrections with only the OPE mechanism, we find $A_{C}=(0.78\pm0.12)\%$, significantly different from zero. However, after including NNLO corrections, the asymmetry reduces to $A_{C}=(0.09\pm0.14)\%$, consistent with a symmetric distribution for “Born-like” events. This indicates that the initial asymmetry arises from higher-order processes, such as IFI and TPE. The measured $A_{C}$ in $\mu^{+}\mu^{-}$ is smaller than the $A_{C} = (3.79 \pm 0.53)\%$ observed for $e^{+}e^{-} \to p\bar{p}$ at $\sqrt{s}=2.125$~GeV, supporting the interpretation that the larger asymmetry in the proton channel is primarily due to OPE$\otimes$TPE contributions.

\section{Summary}
In this study, we report the first clear observation of TPE effects in the TL region, based on high-statistics measurements of the process $e^{+}e^{-} \to p\bar{p}$ with the BESIII detector. 
A statistically significant angular asymmetry ($8.6\sigma$) is observed, which is attributed to TPE contributions, after carefully excluding events with IFI. The extracted asymmetry, as well as the most precise measurements to date of the proton EMFFs in the TL region ($2.000 - 3.671$~GeV), provide new insights into the internal electromagnetic structure and dynamics of the proton.
Our analysis demonstrates that the observed asymmetry is consistent with theoretical expectations from OPE$\otimes$TPE interference and is further validated by comparison with the process $e^{+}e^{-}\to\mu^{+}\mu^{-}$. 
Additionally, the results highlight the impact of TPE corrections on the electric and magnetic form factors ($G_{E}$ and $G_{M}$), and the role of the imaginary part of higher-order form factors, such as $\mathcal{F}_{3}$, in shaping the charge asymmetry.
These findings represent a significant step forward in our understanding of proton structure in the TL region and provide valuable experimental input for future theoretical and phenomenological studies.

\section{Acknowledgments}
This work is supported in part by National Natural Science Foundation of China (NSFC) under Contracts No. 12005219, No. 12375072,  China Postdoctoral Science Foundation No. 2021M693097.


\begin{thebibliography}{99}

\bibitem{Stern1} I.~Estermann, O.~C.~Simpson, and O.~Stern, \href{https://journals.aps.org/pr/abstract/10.1103/PhysRev.52.535}{\color{blue}{Phys. Rev {\bf 52}, 535 (1937)}}.
\bibitem{Hofstadter1} R.~Hofstadter, and R.~W.~McAllister, \href{https://journals.aps.org/pr/abstract/10.1103/PhysRev.98.217}{\color{blue}{Phys. Rev {\bf 98}, 217 (1955)}}.
\bibitem{McAllister1} R.~W.~McAllister, and R.~Hofstadter, \href{https://journals.aps.org/pr/abstract/10.1103/PhysRev.102.851}{\color{blue}{Phys. Rev {\bf 102}, 851 (1956)}}.
\bibitem{Bloom1} E.~D.~Bloom {\it{et al.}}, \href{https://journals.aps.org/prl/abstract/10.1103/PhysRevLett.23.930}{\color{blue}{Phys. Rev. Lett. {\bf 23}, 930 (1969)}}.
\bibitem{Breidenbach1} M.~Breidenbach {\it{et al.}}, \href{https://journals.aps.org/prl/abstract/10.1103/PhysRevLett.23.935}{\color{blue}{Phys. Rev. Lett. {\bf 23}, 935 (1969)}}.
\bibitem{Zhou1} H.~Q.~Zhou, \href{https://journals.aps.org/prc/abstract/10.1103/PhysRevC.95.025203}{\color{blue}{Phys. Rev. C {\bf 95}, 025203 (2017)}}.
\bibitem{VEPP-3_1} J.~Arrington {\it{et al.}}, \href{https://arxiv.org/abs/nucl-ex/0408020}{\color{blue}{arXiv: 0408020 (2004)}}.
\bibitem{VEPP-3_2} I.~A.~Rachek {\it{et al.}}, \href{https://journals.aps.org/prl/abstract/10.1103/PhysRevLett.114.062005}{\color{blue}{Phys. Rev. Lett. {\bf 114}, 062005 (2015)}}.
\bibitem{OLYMPUS1} B.~S.~Henderson {\it{et al.}} (OLYMPUS Collaboration), \href{https://journals.aps.org/prl/abstract/10.1103/PhysRevLett.118.092501}{\color{blue}{Phys. Rev. Lett. {\bf 118}, 092501 (2017)}}.
\bibitem{CLAS1} D.~Adikaram {\it{et al.}} (CLAS Collaboration), \href{https://journals.aps.org/prl/abstract/10.1103/PhysRevLett.114.062003}{\color{blue}{Phys. Rev. Lett. {\bf 114}, 062003 (2015)}}.
\bibitem{CLAS2} D.~Rimal {\it{et al.}} (CLAS Collaboration), \href{https://journals.aps.org/prc/abstract/10.1103/PhysRevC.95.065201}{\color{blue}{Phys. Rev. C {\bf 95}, 065201 (2017)}}.
\bibitem{JLAB1} A.~J.~R.~Puckett {\it{et al.}}, \href{https://journals.aps.org/prl/abstract/10.1103/PhysRevLett.104.242301}{\color{blue}{Phys. Rev. Lett. {\bf 103}, 242301 (2010)}}; \href{https://journals.aps.org/prc/abstract/10.1103/PhysRevC.85.045203}{\color{blue}{{\bf 85}, 045203 (2012)}}.
\bibitem{MAMI1} B.~X.~Gou {\it{et al.}}, \href{https://journals.aps.org/prl/abstract/10.1103/PhysRevLett.124.122003}{\color{blue}{Phys. Rev. Lett. {\bf 124}, 122003 (2020)}}.
\bibitem{Carlson1} C.~E.~Carlson and M.~Vanderhaeghen, \href{https://www.annualreviews.org/doi/10.1146/annurev.nucl.57.090506.123116}{\color{blue}{Ann. Rev. Nucl. Part. Sci. {\bf 57}, 171-204 (2007)}}.
\bibitem{Arrington1} J.~Arrington, P.~G.~Blunden, and W.~Melnitchouk, \href{https://www.sciencedirect.com/science/article/pii/S0146641011000962?via\%3Dihub}{\color{blue}{Prog. Part. Nucl. Phys. {\bf 66}, 782-833 (2011)}}.
\bibitem{Hand1} L.~N.~Hand, D.~G.~Miller, and R.~Wilson, \href{https://journals.aps.org/rmp/abstract/10.1103/RevModPhys.35.335}{\color{blue}{Rev. Mod. Phys. {\bf 35}, 335 (1963)}}.
\bibitem{Rinaldo1} R.~ B.~Ferroli {\it{et al.}}, \href{https://link.springer.com/article/10.1140/epja/i2008-10716-1}{\color{blue}{Eur. Phys. J. A {\bf 39}, 315-321 (2009)}}.
\bibitem{Gakh} G.~I.~Gakh, and E.~Tomasi-Gustafsson, \href{https://www.sciencedirect.com/science/article/abs/pii/S0375947405009905}{\color{blue}{Nucl. Phys. A {\bf 761}, 120-131 (2005)}}.
\bibitem{dianyong} D.~Y.~Chen, H.~Q.~Zhou, and Y.~B.~Dong, \href{https://journals.aps.org/prc/abstract/10.1103/PhysRevC.78.045208}{\color{blue}{Phys. Rev. C {\bf 78}, 045208 (2008)}}.
\bibitem{BABAR} J.~P.~Lee {\it{et al.}} ({\sl{BABAR}} Collaboration), \href{https://journals.aps.org/prd/abstract/10.1103/PhysRevD.87.092005}{\color{blue}{Phys. Rev. D {\bf 87}, 092005 (2013)}}.
\bibitem{Pacetti2} S.~Pacetti, R.~ B.~Ferroli, and E.~Tomasi-Gustafsson, \href{https://www.sciencedirect.com/science/article/pii/S0370157314003184}{\color{blue}{Phys. Rep {\bf 550-551}, 1-103 (2015)}}.
\bibitem{phokhara} H.~Czy{\.z}, J.~ H.~K{\"u}hn, and S.~Tracz, \href{https://journals.aps.org/prd/abstract/10.1103/PhysRevD.90.114021}{\color{blue}{Phys. Rev. D {\bf 90}, 114021 (2014)}}.
\bibitem{Lei1} M.~Ablikim {\it{et al.}} (BESIII Collaboration), \href{https://journals.aps.org/prl/abstract/10.1103/PhysRevLett.124.042001}{\color{blue}{Phys. Rev. Lett. {\bf 124}, 042001 (2020)}}.
\bibitem{Lei2} L.~Xia {\it{et al.}}, \href{https://www.mdpi.com/2073-8994/14/2/231}{\color{blue}{Symmertry {\bf 14}, 231 (2022)}}.

\end{thebibliography}
\end{document}